\documentclass{article}

\usepackage{arxiv}

\usepackage[utf8]{inputenc} 
\usepackage[T1]{fontenc}    
\usepackage{hyperref}       
\usepackage{url}            
\usepackage{booktabs}       
\usepackage{amsmath}
\usepackage{float}

\usepackage{amsfonts}       
\usepackage{nicefrac}       
\usepackage{microtype}      
\usepackage{lipsum}		
\usepackage{graphicx}
\usepackage[authoryear]{natbib}			
\usepackage[nameinlink, capitalize]{cleveref} 
\usepackage{etoolbox} 
\usepackage{epstopdf}

\usepackage{rotating}
\usepackage{pdflscape} 

\usepackage{caption}
\usepackage[flushleft]{threeparttable}
\usepackage{tabularx}
\usepackage{multirow}

\newcommand{\matr}[1]{\mathbf{#1}}

\usepackage{mathtools}
\DeclarePairedDelimiterX{\norm}[1]{\lVert}{\rVert}{#1}

\makeatletter

\patchcmd{\NAT@test}{\else \NAT@nm}{\else \NAT@nmfmt{\NAT@nm}}{}{}

\DeclareRobustCommand\citepos
  {\begingroup
   \let\NAT@nmfmt\NAT@posfmt
   \NAT@swafalse\let\NAT@ctype\z@\NAT@partrue
   \@ifstar{\NAT@fulltrue\NAT@citetp}{\NAT@fullfalse\NAT@citetp}}

\let\NAT@orig@nmfmt\NAT@nmfmt
\def\NAT@posfmt#1{\NAT@orig@nmfmt{#1's}}

\makeatother

\title{Sovereign Default Risk and Credit Supply: Evidence from the Euro Area}

\date{May 10, 2020}	

\author{ \href{https://orcid.org/0000-0000-0000-0000}{Olli Palmén}\\
	Faculty of Social Sciences\\
	University of Helsinki\\
	P.O. Box 17 (Arkadiankatu 7), \\
	00014 University of Helsinki, Finland \\
	\texttt{olli.palmen@helsinki.fi} \\
}

\renewcommand{\headeright}{Working Paper}
\renewcommand{\undertitle}{Working Paper}

\hypersetup{
pdftitle={Sovereign Default Risk and Credit Supply},
pdfsubject={},
pdfauthor={Olli Palmén},
pdfkeywords={Sovereign risk shocks, euro area, structural vector autoregression},
}

\begin{document}
\maketitle

\begin{abstract}
Did sovereign default risk affect macroeconomic activity through firms' access to credit during the European sovereign debt crisis? We investigate this question by a estimating a structural panel vector autoregressive model for Italy, Spain, Portugal, and Ireland, where the sovereign risk shock is identified using sign restrictions. The results suggest that decline in the creditworthiness of the sovereign contributed to a fall in private lending and economic activity in several euro-area countries by reducing the value of banks' assets and crowding out private lending.
\end{abstract}

\keywords{Sovereign debt crisis \and  credit supply \and structural vector autoregression}

\section{Introduction} \label{sec:introduction}


The link between bank and sovereign credit risk is often considered as a key transmission mechanism of the European sovereign debt crisis. Starting in 2009, concerns over the sustainability of government debt in several peripheral European countries led to a sudden increase in banks' credit risk, which was largely due to banks' exposure to sovereign debt exposures. The resulting decrease in asset values led to a tightening in banks' financing conditions and the subsequent contraction in bank lending (see, e.g., \cite*{Acharya2015}, \cite{Bocola2016}, and \cite{Brunnermeier2016}). The negative feedback loop between the sovereign and the real economy reinforced the recessionary impact of the credit crunch; weak economic activity further strained the fiscal position of the sovereign and put upward pressure on sovereign borrowing costs \citep{Brunnermeier2016}.

Banks' exposure to sovereign debt transmits sovereign risk to the real economy in three ways. First, a decline in the value of banks' assets restricts the banks' access to funding, which leads to a reduction in private lending \citep*{Bocola2016, Gennaioli2018}. Second, a decline in sovereign bond prices may tighten banks' funding conditions, given that government debt is often used as collateral in the interbank lending \citep{Engler2016}. Third, banks may reduce private lending by changing the compositions of their balance sheets during a crisis \citep*{Acharya2018, Becker2017}.

Although the effect of sovereign risk shocks on bank lending has garnered much attention in recent years (see, e.g., \cite{Acharya2015}, \cite{Bahaj2019}, \cite*{Bofondi2017}, and \cite{Popov2014}), the extent to which sovereign default risk accounts for the decrease in credit supply and the subsequent decrease in economic activity is not clear. In this study, we re-examine the credit channel of sovereign default risk and provide cross-country evidence about its effect on real economic activity during the European sovereign debt crisis. We proceed by estimating a structural panel vector autoregressive (SVAR) model for Italy, Spain, Portugal, and Ireland, using monthly data covering the period 2003:M1-2018:M12. We identify a sovereign risk shock using sign restrictions that disentangle the origin of the shock from other shocks that affect the supply of credit. The use of macroeconomic data provides a new perspective on the negative effects of the sovereign risk shock on bank lending, which has been previously mainly examined using bank-level data. In comparison to studies using micro-level data, the panel SVAR approach helps to describe general equilibrium or substitution effects and is able to capture potential feedback effects between sovereign risk and economic activity.

The results indicate that the macroeconomic effects of the sovereign risk shock varied across the crisis-hit countries during the sovereign debt crisis. A contractionary sovereign risk shock appears to have a large and persistent negative effect on bank lending and real economic activity. Moreover, the sovereign risk shock partly explains the decline in economic output during the sovereign debt crisis in Italy, Portugal and Spain. The findings also imply that the sovereign risk shock led to an increase in economic output prior to the crisis. Moreover, the protracted increase in the home bias of banks' government debt holdings also suggests that banks' portfolio reallocation from private lending to sovereign debt might have cut back the credit supply by crowding out lending. The results are also important from a policy perspective, given that the policies to prevent the spillover from sovereign risk to banks would seem to be effective in mitigating or preventing the crisis. Moreover, the fact that banks' holding of sovereign debt largely contributed to the crisis implies the need for a European safe-asset as a store of value, as proposed by \cite{Brunnermeier2016}.

The rest of the paper is organized as follows. \cref{sec:literature} reviews the related literature. \cref{sec:methods} provides an overview of the methodology, the identification, and the data.  \cref{sec:results} presents and discusses results. \cref{sec:conclusion} concludes.

\section{Literature review} \label{sec:literature}

Previous studies document a strong link between the yields on government debt and business cycles in developing and, more recently, in developed economies. \cite{Uribe2006} find that international financial shocks explain a large share of the business cycle variation in emerging countries through their effect on domestic borrowing costs. They also find that the response of domestic interest rate spreads to domestic fundamentals further reinforces the shocks. \cite{Neri2015} use a large-dimensional factor augmented vector autoregression (FAVAR) model to identify macroeconomic effects of increases in sovereign risk during the European sovereign debt crisis. They show that sovereign debt tensions had a sizeable effect on economic activity by tightening bank-lending conditions. \cite{Bahaj2019} uses high-frequency government bond yield movements around key events during the European sovereign debt crisis to study the macroeconomic impact of innovations to sovereign risk premia in Spain, Ireland, Italy and Portugal in a Bayesian panel SVAR. The results suggest that unanticipated changes in government bond risk premia can explain approximately a third of the variation in the unemployment rate and a sizeable share of the variation in borrowing costs, which lends support to the significant role of the banking sector in the transmission of the crisis. Although previous research documents a large role of the banking sector, the transmission mechanism is not explicit. This paper contributes to the existing literature by examining the role of the banking sector in the transmission of sovereign risk to the real economy.

This paper is also related to previous studies that use sign-identified SVAR models to study the macroeconomic effects of financial shocks. It deviates from previous research by distinguishing the financial shocks originating from unanticipated changes sovereign risk\footnote{The method we use to determine the origin of financial shocks resembles the identification strategy in \cite*{Furlanetto2017} based on sign restrictions in a structural VAR model to decompose financial shocks into housing market, credit market, and uncertainty shocks.}. \cite{Gambetti2016} study the effects of loan supply shocks on the business cycle in the euro area, the UK and the US by estimating a VAR model with time-varying parameters and stochastic volatility. They find that loan supply shocks play a significant role in the variation of economic activity, especially during recessions. \cite{Bijsterbosch2015} find that a credit supply shock spurred economic growth before the financial crisis but worsened the recession during the crisis in all countries. However, from the beginning of the European sovereign debt crisis, the credit supply shock contributed negatively to the growth of output in crisis-hit countries, whereas it affected economic growth positively in the core countries. \cite*{Hristov2012} employ a panel SVAR for eleven euro-area member countries and find that credit supply shocks significantly affected real gross domestic product and loans issued by banks in all countries during the financial crisis.

This paper is also closely related to studies using micro-level data to examine the effect of sovereign debt crisis on bank loan volumes (see, e.g., \cite{Acharya2015}, \cite{Acharya2018}, \cite{Bofondi2017}, \cite{DeMarco2019}, and \cite{Popov2014}). \cite{Acharya2018} study the effects of the sovereign debt crisis on bank lending using data syndicated loans for European non-financial corporations. They find that the reduction in bank lending to firms in crisis-hit countries was largely explained by a reduction in the value of banks' holdings of sovereign debt and crowding out private borrowing through further government debt purchases. In all, the credit crunch induced by the risk of government default can explain up to one-half of negative real effects during the sovereign debt crisis. \cite{Popov2014} examine the cross-border transmission of sovereign debt tensions by studying the issuance of syndicated loans by banks domiciled in 11 European non-crisis countries during the sovereign debt crisis. They document a significantly smaller increase in loan issuance after 2009 by banks that were exposed to government debt of crisis-hit countries. This paper complements the micro-level evidence by taking into account both macroeconomic as well as general equilibrium effects of financial shocks.

\section{The model and the identification strategy} \label{sec:methods}

This section provides an overview of the econometric model, the data, and the identification approach.

\subsection{The panel SVAR model}

We estimate a Bayesian panel SVAR model for Italy, Ireland, Portugal and Spain; the setup is similar to that of \cite{Jarocinski2010}, who used it to study the effects of monetary policy. The partially pooled model is chosen to allow for cross-country heterogeneity in the dynamics of the model, while retaining similarities between the economic structures in each country. Moreover, in comparison to individually estimated models, the use of panel data is likely to improve the quality of the estimates of the model. However, the expected increase in efficiency comes at the cost of losing information in the cross-sectional dimension by shrinking the estimates around a common mean. The model is identified using sign and exclusion restrictions on the impulse response functions by using the algorithm of \cite*{Arias2018}\footnote{The estimation of the model is done in Matlab using the Bayesian Estimation, Analysis and Regression (BEAR) toolbox \citep*{BEAR}.}. 

For each country $c$, the reduced-form VAR model model is

\begin{equation}
\matr{y}_{c,t} = \matr{\Gamma}_{c}\matr{z}_{c,t} + \sum_{l=1}^{L} \matr{B}_{c,l} \matr{y}_{c,t-l} + \matr{u}_{c,t},
\label{var}
\end{equation}

where $\matr{y}_{c,t}$ is an $N \times 1$ vector of endogenous variables. $\matr{\Gamma}_{c}$ is an $N \times M$ coefficient matrix for the $M \times 1$ vector of deterministic variables $\matr{z}_{c,t}$, $\matr{y}_{c,t-l}$ is the $l$-th lag of vector of endogenous variables, $\matr{B}_{c,l}$ are the $N \times N$ country-specific coefficient matrices for the $l$-th lag of endogenous variables, and $\matr{u}_{c,t}$ is a $N \times 1$ vector of Gaussian white noise error terms with $\matr{u}_{c,t} \sim \mathcal{N}(\matr{0},\matr{\Sigma}_c)$. 

For each country, the model may be written in matrix notation by stacking $\matr{y}_{c,t}'$ for all observations $t$ such that

\begin{equation}
\matr{Y}_{c} = \matr{X}_{c}\matr{B}_{c} + \matr{Z}_{c}\matr{\Gamma}_{c} + \matr{U}_{c},
\label{matrix_var}
\end{equation}

where $\matr{Y}_{c} \equiv [\matr{y}_{c,1}' , \ldots, \matr{y}_{c,T}']'$, $\matr{X}_{c} \equiv [\matr{X}_{c,0}', \ldots, \matr{X}_{c,T-1}']'$,  with $\matr{X}_{c,t} \equiv [\matr{y}_{c,t-1}', \ldots, \matr{y}_{c,t-L}']$ and $\matr{B}_{c} \equiv [\matr{B}_{c,1}', \ldots \matr{B}_{c,L}']' $.

The likelihood for each country specified in \cref{matrix_var} may be expressed in vectorized form as

\begin{equation}
p\left(\matr{y}_c \vert \matr{\beta}_c, \matr{\gamma}_c, \matr{\Sigma}_c\right) = \mathcal{N}\left(\left( \matr{I}_n \otimes \matr{x}_c \right) \matr{\beta}_c + \left( \matr{I}_m \otimes \matr{z}_c \right) \matr{\gamma}_c , \  \matr{\Sigma}_c \right),
\end{equation}

where $\matr{y}_c \equiv \text{vec}\left({\matr{Y}_c}\right)$, $\matr{X}_c \equiv \text{vec}\left({\matr{X}_c}\right)$, $\matr{\beta}_c \equiv \text{vec}\left({\matr{B}_c}\right)$, $\matr{z}_c \equiv \text{vec}\left({\matr{Z}_c}\right)$, $\matr{\gamma}_c \equiv \text{vec}\left({\matr{\Gamma}_c}\right)$, and $\matr{u}_c \equiv \text{vec}\left({\matr{U}_c}\right)$.

The prior for the coefficient of deterministic variables is assumed to be non-informative such that $\matr{\gamma} \propto 1$. We also assume a diffuse prior for the error variances $\matr{\Sigma}_c$ of the form

\begin{equation}
p(\matr{\Sigma}_c) \propto \vert \matr{\Sigma}_c \vert^{-\frac{1}{2}\left(N+1 \right)}.
\end{equation}

We specify an exchangeable prior distribution for the coefficients of the endogenous variables $\beta_{c}$, which has common mean and variance across countries:

\begin{equation}
p(\matr{\beta}_c \vert \matr{b}, \matr{\Lambda})  \sim \mathcal{N}( \matr{b} ,\ \matr{\Lambda}).
\end{equation}

The prior for the mean is specified to be non-informative $\matr{b} \propto 1$. The covariance matrix $\matr{\Lambda}$ has a Minnesota-type prior $\matr{\Lambda} \equiv \lambda_1 \matr{\Omega}$, for which the off-diagonal elements of $\matr{\Omega}$ are zero and the diagonal elements may be positive and non-zero. For each lag $l$, equation $i$ and variable $k=1,\ldots,N$, the standard deviation of the coefficient is $1 / l^{\lambda_3}$ if $i=j$ and $\sigma_i \lambda_2/\sigma_j$ if $i \neq j$. The ratio $\sigma_i/\sigma_j$ is included to account for the different magnitude of the coefficients. Because the variance is assumed to be the same for each country, in practice, $\sigma_i$ is calculated as the standard deviation of the pooled regression for each variable on $L$ lagged values.

Following \cite{Jarocinski2010}, we set $\lambda_2 = 1 $ and $\lambda_3 = 0$. The overall tightness of the prior is determined by a parameter $\lambda_1$. As $\lambda_1$ approaches zero the coefficients shrink toward the common mean $\matr{b}$, which results in a pooled model by reducing country-specific variation. As $\lambda_1$ grows, the more the country-specific coefficients are allowed to vary. Therefore, $\lambda_1$ essentially determines the extent to which the dynamics of the model vary across countries. In practice, $\matr{\Omega}$ is treated as fixed, but the overall tightness of the prior $\lambda_1$ is considered to be random. We specify a weakly informative inverse-Gamma prior for $\lambda_1$, such that 

\begin{equation}
p\left( \lambda_1 \vert s, \ \nu \right) \propto \lambda_1^{-(s+1)} \text{exp}\left(-\frac{\nu}{\lambda_1}\right).
\end{equation}

Following \cite{BEAR}, we set $s=0.001$ and $\nu=0.001$ to obtain a prior of the form $p(\lambda) = \lambda_1^{-\frac{1}{2}}$. The model is estimated using a Gibbs sampler, as detailed in \cite{Jarocinski2010}\footnote{The results are robust to a range of other weakly informative priors between 0.1 and 0.001, which have been used by \cite{Gelman2006} and \cite{Jarocinski2010}.}.

\subsection{Data}

The panel VAR model is estimated for Italy, Spain, Ireland, and Portugal using monthly observations, covering the period from 2003:1 to 2018:12\footnote{The start of the sample is dictated by the availability of the balance-sheet data of the monetary and financial institutions provided by the European Central Bank (ECB).}\footnote{We exclude Greece from the analysis due to the break in the bond yield series following the debt restructuring, although it was one of the countries that was most severely affected by the sovereign debt crisis.}. The data includes the following variables for each country: a measure of real output, an index of consumer prices, volume of loans to non-financial firms, a composite retail lending rate, and a measure of the home bias of domestic banks' holdings of government debt. In addition, a shadow short rate for the euro area is also included to indicate the monetary policy stance. The model thus includes variables that cover the financial and real sectors of the economy and which have been used in previous literature to identify shocks to the banking sector (see, e.g., \cite{Gambetti2016}, \cite{Hristov2012}, and \cite{Peersman2011b}).

Industrial output is used as a measure of real economic output in the baseline model. Industrial output enters as the log of a wide production index. The volume of loans is measured by the log of total loans to non-financial companies adjusted for loan sales and securitization. The government bond spread is defined as the difference between the benchmark 10-year yield and the corresponding 10-year zero-coupon swap rate, which is used as a measure of the risk-free rate. The home bias of banks' holdings of government debt is calculated as the ratio of domestic government bonds to foreign government debt held by the monetary and financial institutions in the domestic country, excluding the European System of Central Banks. A shadow short rate by \cite{Krippner2013} is included as a measure monetary policy stance for the euro area. A detailed description of the data and its sources is presented in \Cref{app:data}.

\subsection{Identification of the sovereign risk shock in the baseline model} \label{sec:identification}

We use sign restrictions on the impulse responses to identify a sovereign risk shock. Sign restrictions are commonly used in applied macroeconomic literature to identify economic shocks, as pioneered by \cite{Faust1998}, \cite{Canova2002}, \cite{Uhlig2005}, and \cite{Peersman2005}, and more recently, sign restrictions have been used to identify credit supply and financial shocks (see, e.g., \cite{Hristov2012}, \cite{Gambetti2016}, and \cite{Furlanetto2017})\footnote{In their critique of sign-identified structural VAR models, \cite{Baumeister2015} caution that uninformative priors may be inadvertently informative in sign-identified structural VAR models. To alleviate some concerns about uninformative priors influencing the estimation, a number of robustness checks were conducted: alternative endogenous variables, a shorter sample length and alternative restrictions on the timing of effect of the shocks on the variables. While in principle identifying the importance of priors in influencing the results may have benefits, there is disagreement as to whether laying out explicit prior beliefs is relevant in practice \cite{Kilian2019}.
}. 

In comparison to a recursive structural model, sign restrictions require less stringent assumptions about the contemporaneous relationships between the variables. For example imposing exclusion restrictions on the impact multiplier matrix for financial variables may not be justifiable from a theoretical perspective, given that these variables respond quickly to new information (see discussion in \cite{Peersman2005} and \cite{Bjornland2009}.

We proceed by identifying three different shocks: a sovereign risk shock, a credit supply shock, and a credit demand shock. The structural shocks are identified using a combination of sign and zero restrictions using the algorithm by \cite{Arias2018}. The details of the algorithm are provided in \Cref{algorithm}. The restrictions are summarized in \Cref{tab:restrictions}.

\begin{table}[h!]
\small
  \begin{threeparttable}
  \centering
    \caption{Sign restrictions in the baseline model}
      \label{tab:restrictions}
      \renewcommand{\arraystretch}{0.9}
     \begin{tabularx}{\linewidth}{lccc}
     \\
        \toprule
              & \multicolumn{3}{c}{\textit{Shock}}\\
          		\textit{Variable} 		& Credit demand shock & Credit supply shock & Sovereign risk shock\\
        \midrule
        Output 		& $-$  & $-$ & $-$	\\
        Prices 			  		& $\bullet$ & $\bullet$ & $\bullet$ \\
        Loan volume 	  		& $-$ & $-$ & $-$	\\
        Retail lending rate & $-$ & $+$ & $+$	\\
        Banks' home bias & $\bullet$  & $0$  & $+$	\\
        Government bond spread 			& $\bullet$  & $\bullet$  & $+$	\\
        Short rate 			& $\bullet$ & $\bullet$  & $-$	\\
        \bottomrule
     \end{tabularx}
       \captionsetup{justification=centering}
    \begin{tablenotes}
      \small
      \item \textit{Notes:} $+$/$-$ denote the sign on impact of impulse responses for each variable to a contractionary shock. The sign is set to $0$ if the response of the shock is restricted to zero on impact. The responses of output, loan volume and the lending rate are restricted for periods three to six following the shock. The response is left unrestricted, if the sign is indicated by $\bullet$.
    \end{tablenotes}
  \end{threeparttable}

\end{table}

The identification of thr credit demand and the credit supply shock is standard in the literature (e.g., \cite{Peersman2011b}, \cite{Gambetti2016} and \cite{Hristov2012}) and is based on general macroeconomic models (see e.g, \cite{Curdia2010}, \cite*{Gerali2010}, \cite*{Gertler2011})\footnote{See \cite{Gambetti2016} and \cite{Hristov2012} for a summary of sign restrictions used in theoretical models.}. 

The credit demand shock causes the volume of loans and the retail lending rate to move in the same direction. A positive credit demand shock increases the volume of loans and simultaneously increases the cost of lending due to inelastic supply of loans. The credit demand shock may be interpreted as an unanticipated change in firms' and households' demand for loans, for example following changes in aggregate demand. Positive co-movement between retail lending rates and loan volumes may also be associated with loosening borrowing constraints.

The credit supply shock is related to exogenous changes in banks' lending to firms and households, for example unanticipated change in banks' capital, funding conditions or new regulations. It also encompasses unanticipated changes in monetary policy, which shows up as a change in the short rate. A positive credit supply shock increases the supply of loans to firms and households and decreases the cost of lending. Therefore, their distinct effects on the co-movement between loan volumes and retail lending rates distinguish the credit supply shock and the credit demand shock from each other.

The sovereign risk shock affects the real economy by affecting the supply of credit. The theoretical literature finds that sovereign default risk affects the financial sector due to banks' large exposure to government debt (see e.g., \cite{Bocola2016}). An increase in sovereign borrowing costs reduces the value of the banks' balance sheets and tightens their funding constraints as well as makes loans to households and firms more risky. This subsequently reduces the amount of loans banks issue to domestic firms and households, resulting in a credit crunch.

To disentangle the credit supply shock and the sovereign risk shock, we introduce additional restrictions on the impulse responses. Namely, we restrict the response of the home bias of banks' holdings of sovereign debt to be positive following a sovereign risk shock, whereas the credit supply shock that is not associated with a change in sovereign risk is not assumed to affect banks' home bias. This assumption is consistent with the observation that banks in distressed countries have increased their holdings of domestic sovereign debt in response to sovereign default risk. \cite*{Battistini2014} study the effects of sovereign default risk on the domestic sovereign exposures of banks in the euro area. They find that banks in peripheral countries increased their domestic exposure to sovereign debt in response to a country-level shock, whereas banks in core countries did not. However, banks in both peripheral and core countries increased their domestic exposure in response to systemic shocks. Their findings support the idea that banks' in peripheral countries increased their exposure for higher returns and in response to 'moral suasion' or pressure by domestic regulators. Moreover, domestic banks have a comparative advantage in bearing risk in the case of a break-up of the euro area, whereby liabilities and assets would be redenominated into new currency. 

According to \cite*{Broner2014} banks increase their holdings of domestic sovereign debt when sovereign default risk increases, because sovereign debt delivers a higher expected return to domestic investors than to foreign investors. Discrimination of foreign investors may arise due to regulations or moral suasion, whereby regulators induce domestic banks to take on risk to support the demand for bonds. Home bias of banks' sovereign debt holdings may also result from high risk trades, or so-called 'carry trades' in which banks used low funding costs to invest in higher yielding sovereign bonds. Particularly weakly-capitalized banks increased their exposure to domestic high yielding sovereign debt, because the bank would profit in all circumstances except when both the government and the bank were in default \citep{Acharya2015}. 

The sovereign risk shock is assumed to affect the home bias of banks' holdings of sovereign debt on impact, such that banks increase their share of domestic government bonds to foreign government bonds as the government bond spread increases. We restrict the response of banks' home bias on impact following a credit supply shock to zero. Home bias of government debt is defined as the ratio of domestic government bonds to foreign government bonds held by the domestic monetary and financial institutions, excluding the Eurosystem. We also restrict the response of the short rate to be negative following a contractionary sovereign risk shock to account for the negative impact on output, given that a country-specific specific sovereign risk shock poses a systematic risk to the whole monetary union\footnote{The results are not changed when the sign of the response of the short rate is left unrestricted.}. This is also consistent with the assumption that banks increase their domestic government bond holdings by exploiting lower borrowing costs.

We remain agnostic about the sign of the effect on the short rate of the credit supply and credit demand shocks, due to the assumption that the European Central Bank considers the overall economic conditions in the whole euro area in making monetary policy decisions instead of the conditions of a single member state. All identified structural shocks are restricted to affect output, lending to firms and household and the retail lending rate with the given signs three to six months following the shock, while the other restrictions only apply on impact\footnote{The results are robust to restrictions imposed only on impact. Similar timing restrictions, which impose a lagged response of real economic variables to shocks originating in the financial sector, have been used in previous empirical studies (see e.g., \cite{Peersman2011b} and \cite{Barnett2014}). Restricting the response of output is necessary for identification.} Moreover, we leave the sign of inflation unrestricted due to uncertainty about whether aggregate supply or aggregate demand effects dominate following shocks to credit (see discussion in \cite*{Abbate2016}).

A particular concern with disentangling the credit supply and sovereign risk shocks is related to the negative feedback loop between economic activity and government bond spreads. Independent of their origin, credit supply shocks have a recessionary impact on economic output, which may further deteriorate the fiscal position of the sovereign and limit the banks' ability to lend to the private sector \citep{Brunnermeier2016}. Moreover, a credit supply shock may be interpreted as a sovereign risk shock if the credit supply shock increases government bond spreads and banks' home bias of government debt within the period. However, given that we estimate the model using monthly data, it is reasonable to assume that it takes over one period for the feedback loop to complete.

\section{Results} \label{sec:results}

This section presents the results of the partially pooled panel VAR. In addition to country-specific estimates, we also present results for the fully pooled model for comparison. Each model is estimated using four lags for each country. The suggested lag length based on the Bayesian information criterion (BIC) ranges between one and two for individual countries. Given that the data is monthly, we account for possible seasonality in the data by including further lags. The results are based on 110000 draws from the Gibbs sampler with 10000 initial draws discarded as burn-in\footnote{The baseline results are robust to a shorter sample period, a range of alternative priors and alternative measures of monetary policy. To save space, the results are not reported in detail, but they are available on request.}.

\subsection{Real effects of sovereign risk shocks}

\begin{figure}
\caption{Impulse response functions to a contractionary sovereign risk shock in the pooled panel SVAR model}
\bigskip
\includegraphics[scale=0.9]{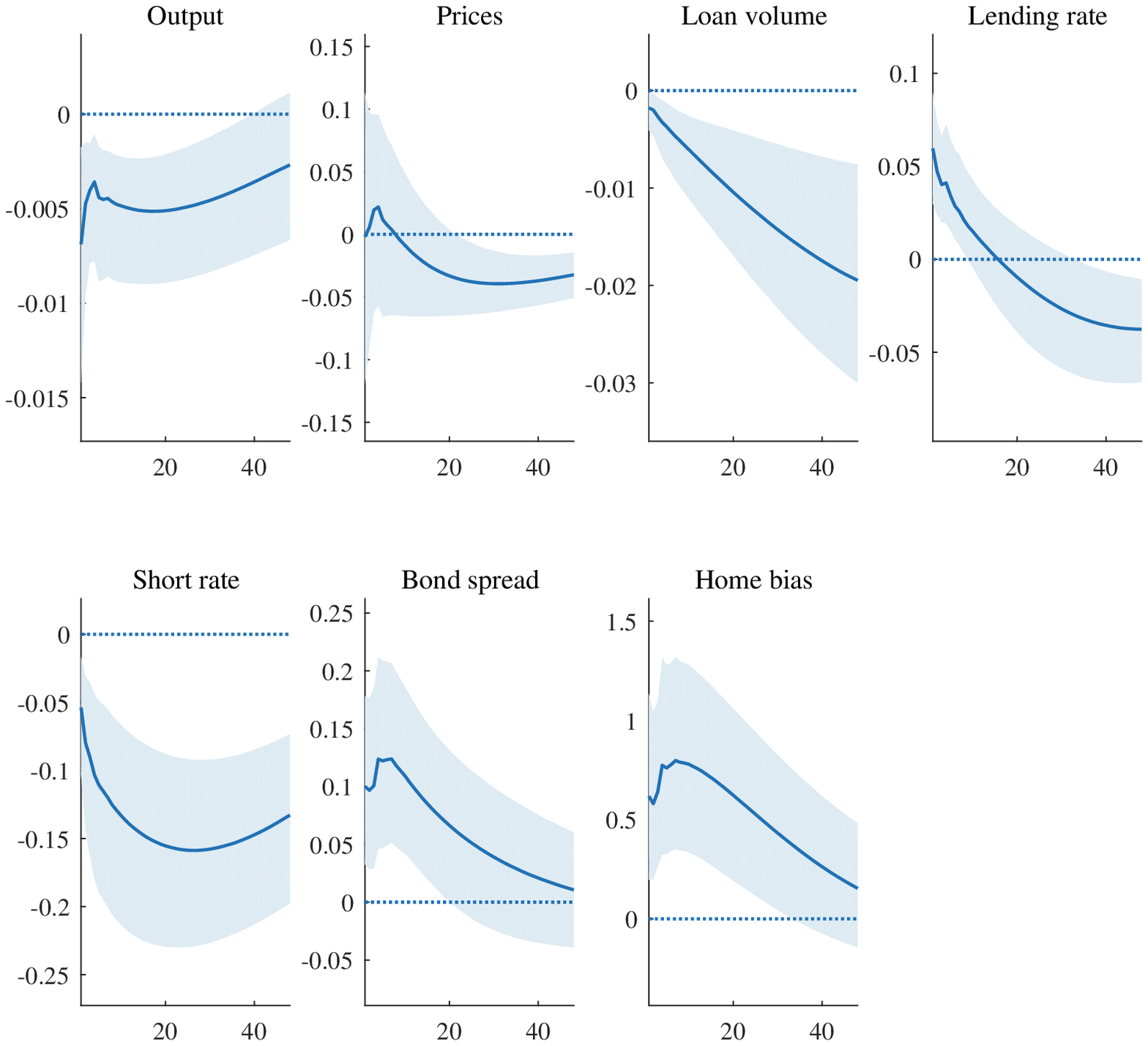}
\smallskip
\caption*{\textit{Notes}: The solid line is the posterior median and the shaded regions mark the 68 \% credible intervals. The shock is normalized, such that the the government bond spread increases by 10 basis points on impact.}
\label{irf_sr_pooled}
\end{figure}

Impulse response functions are used to illustrate the macroeconomic impact of the structural shocks. \Cref{irf_sr_pooled} traces out the effect of a contractionary sovereign risk shock on the endogenous variables in the pooled SVAR model. The shock is normalized such that the median response of the government bond spread is 10 basis points on impact.

A contractionary sovereign risk shock leads to an immediate decrease in output and the permanent fall in loan volumes. The effect on output is negative with high posterior probability for almost 40 months following the shock. There is also a persistent and negative effect on loan volumes. The shock also has an immediate positive effect on the retail lending rate. A 10 basis point increase in the government bond spread is associated with a median increase in the retail lending rate of five basis points. The size of the response is close to the effect reported by \cite{Bahaj2019}, who finds a 30 basis point increase in private sector funding costs with a 100 basis point increases in the 2-year government bond spread. The median response of the retail lending rate turns negative after more than a year following the shock. The short-lived increase and the subsequent fall in the retail lending rate may be caused by the accommodative stance of monetary policy. The short rate falls by 5 basis points on impact and continues to decline for two years following the shock. A contractionary sovereign risk shock also leads to a large and persistent increase in the home bias of banks' holdings of sovereign debt. The effect peaks shortly after the impact and dissipates largely after two years. The response of prices is less precisely estimated, however a contractionary sovereign shock leads to a persistent decline in prices with high probability after an immediate positive median response.

\begin{sidewaysfigure}
\caption{Impulse response functions to a contractionary sovereign risk shock in the partially pooled panel SVAR model}
\bigskip \bigskip
\includegraphics[scale=0.9]{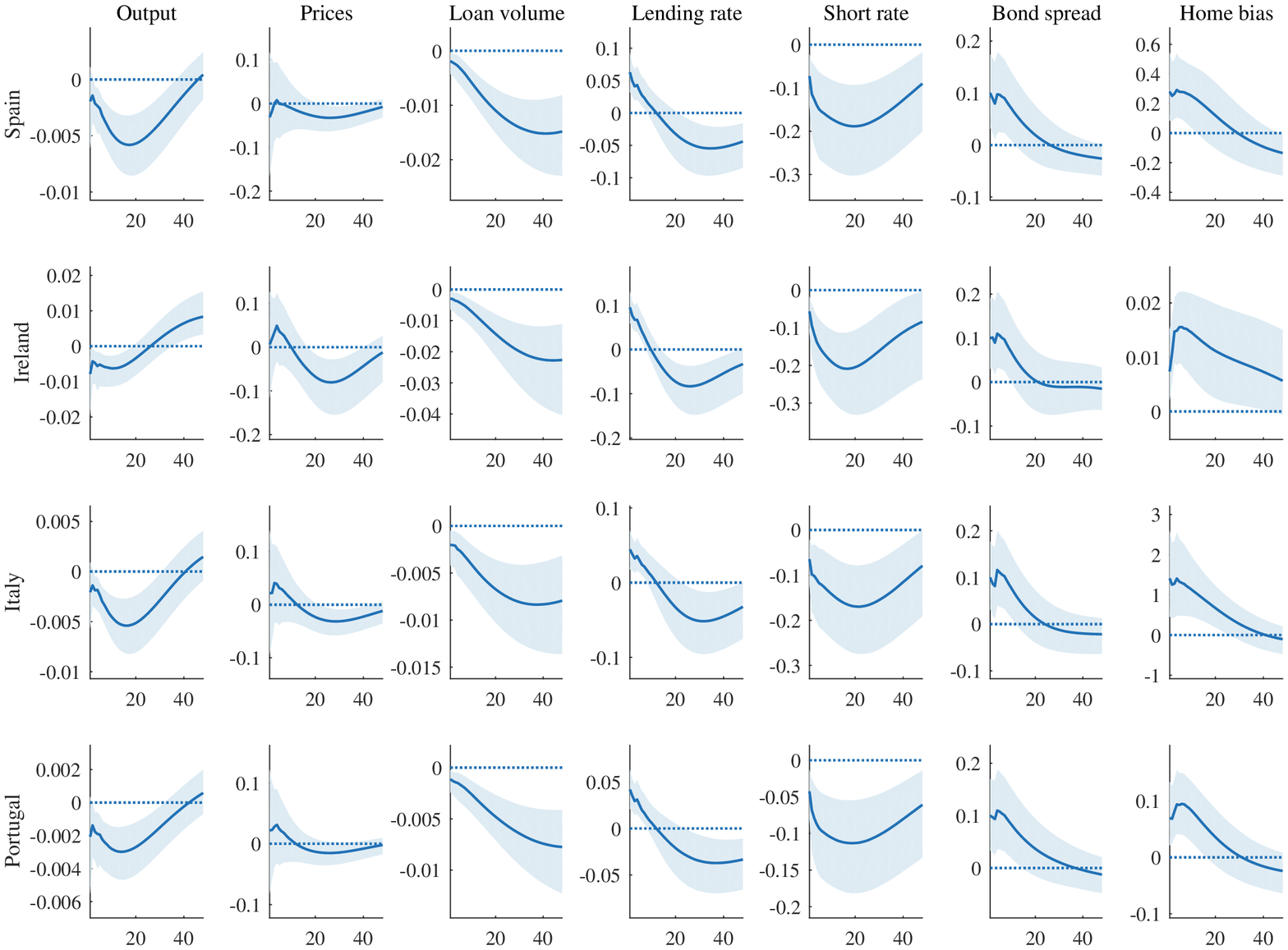}
\smallskip
\caption*{\textit{Notes}: See the notes in \Cref{irf_sr_pooled}}
\label{irf_sr}
\end{sidewaysfigure}

The impulse responses of the country-specific model in \Cref{irf_sr} are generally consistent with the pooled estimates. The impulse response functions show that the impact effects of a contractionary sovereign risk shock on the retail lending rates vary by country. A 10 basis point increase in the government bond spread is associated with an increase in the the median retail lending rate by 5 basis points on impact in Spain, Portugal, and Italy and 10 basis points in Ireland. A sovereign risk shock also results in a large and persistent increase in the home bias of banks' holdings of government debt in Italy and Spain and to a lesser extent in Ireland and Portugal.

Next, we assess the relative importance of the sovereign risk shock on output during the sample period by using historical decompositions to construct a counterfactual time series of each variable that removes the effect of sovereign risk shocks. To that end, the methodology of \cite{Kilian2014} is used.

\begin{figure}
\caption{The difference between the actual output and the counterfactual in the absence of the sovereign risk shock}
\bigskip
\includegraphics[scale=0.9]{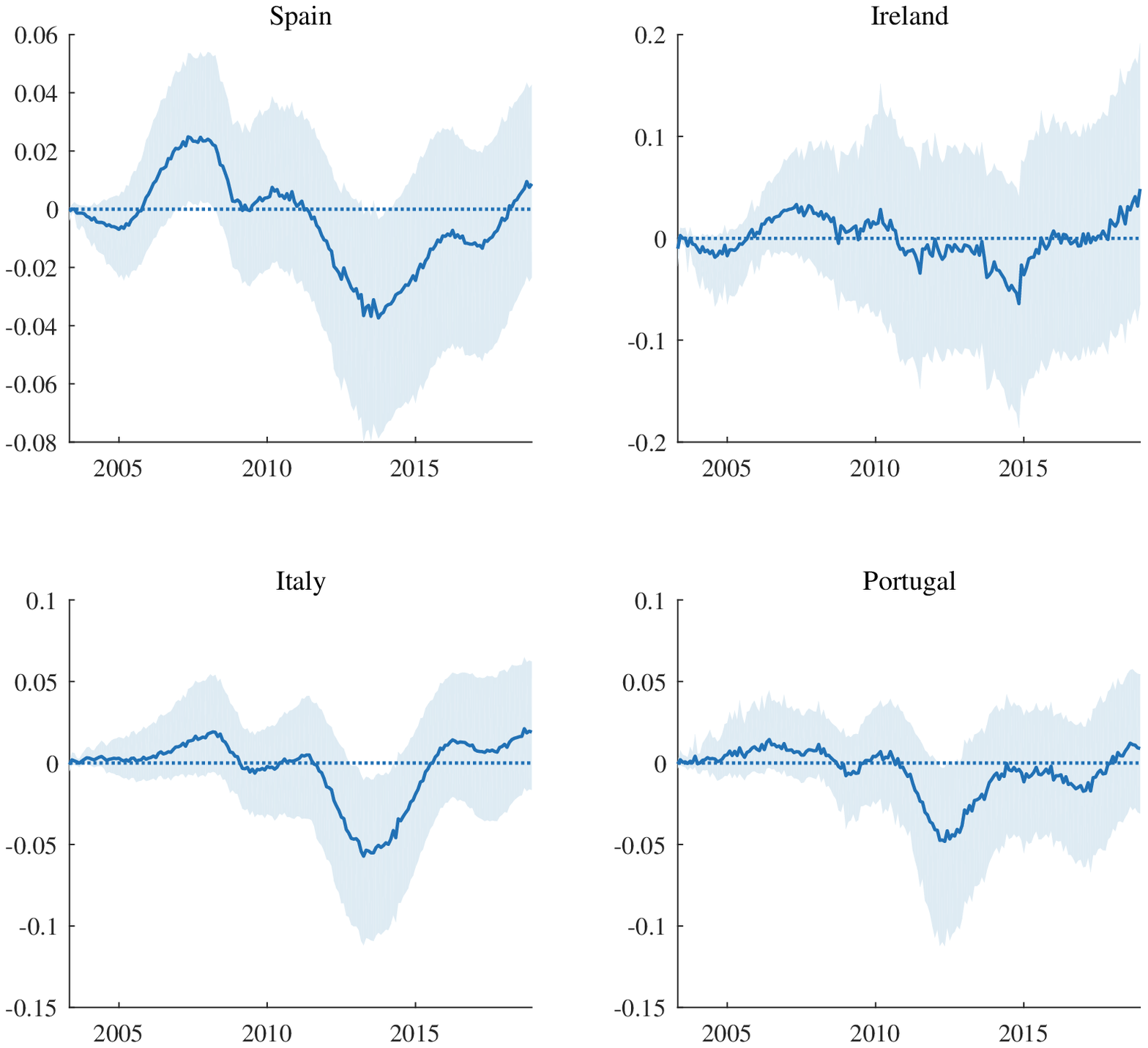}
\smallskip
\caption*{\textit{Notes}: The solid line is the median difference between the actual output and the counterfactual. Shaded areas mark the 68 \% credible intervals.}
\label{cf13}
\end{figure}

The contribution of the sovereign risk shock to output in each country is presented in \Cref{cf13}, which shows the difference between the actual output and the median output in the absence of sovereign risk shocks in each country and the 68 \% credible intervals. The results suggest that sovereign risk shocks contributed to the decline in industrial output in Italy, Portugal and Spain after 2011. Based on the median estimates, at the height of the crisis, output would have been approximately 3-5 \% higher in these countries in the absence of sovereign risk shocks. The effect of sovereign risk shocks on output is negligible in Ireland. However, the estimations that use unemployment as an alternative measure of economic activity show that sovereign risk shocks also contributed to a median increase in unemployment in Ireland, as discussed in \Cref{sec:alt}. In Portugal, the negative effect on output is more short-lived compared to Italy and Spain, which coincides with the countries re-access to financial markets after the government bailout.

\begin{table}
\small
  \begin{threeparttable}
  \centering
    \caption{Forecast error variance decomposition for the sovereign risk shock}
    \bigskip
      \label{tab:fevd}
      \renewcommand{\arraystretch}{1.2}
     \begin{tabularx}{\linewidth}{lp{5mm}ccccp{5mm}cccc}
     \\
        \toprule
     &   & \multicolumn{4}{c}{Spain} & & \multicolumn{4}{c}{Ireland}\\
     &   & \multicolumn{4}{c}{\textit{Horizon}} & & \multicolumn{4}{c}{\textit{Horizon}}\\ 
 
         \textit{Variable} 			&	& 1 & 6 & 12 & 24 & & 1 & 6 & 12 & 24\\
       \cline{1-1} \cline{3-6} \cline{8-11} 
Output&&0.02&0.04&0.09&0.14&&0.01&0.02&0.02&0.03\\
Prices&&0.10&0.09&0.07&0.08&&0.09&0.07&0.05&0.05\\
Loan volume &&0.04&0.06&0.10&0.14&&0.02&0.03&0.04&0.06\\
Retail lending rate&&0.15&0.06&0.03&0.03&&0.19&0.07&0.04&0.04\\
Banks' home bias&&0.06&0.07&0.09&0.10&&0.05&0.09&0.12&0.14\\
Government bond spread&&0.16&0.13&0.12&0.11&&0.09&0.09&0.08&0.07\\
Short rate&&0.09&0.10&0.11&0.11&&0.06&0.21&0.25&0.24\\
\\

     &   & \multicolumn{4}{c}{Italy} & & \multicolumn{4}{c}{Portugal}\\
     &   & \multicolumn{4}{c}{\textit{Horizon}} & & \multicolumn{4}{c}{\textit{Horizon}}\\ 
         \textit{Variable} 		&		& 1 & 6 & 12 & 24 & & 1 & 6 & 12 & 24\\
       \cline{1-1} \cline{3-6} \cline{8-11} 
Output&&0.02&0.04&0.09&0.14&&0.02&0.04&0.07&0.10\\
Prices&&0.09&0.09&0.08&0.08&&0.10&0.10&0.09&0.09\\
Loan volume &&0.05&0.05&0.07&0.10&&0.05&0.05&0.07&0.10\\
Retail lending rate&&0.16&0.07&0.04&0.04&&0.15&0.07&0.04&0.04\\
Banks' home bias&&0.05&0.05&0.07&0.08&&0.07&0.10&0.12&0.13\\
Government bond spread&&0.14&0.14&0.13&0.11&&0.18&0.17&0.16&0.14\\
Short rate&&0.11&0.11&0.12&0.13&&0.09&0.15&0.18&0.17\\

\bottomrule
     \end{tabularx}
       \captionsetup{justification=centering}
    \begin{tablenotes}
      \small
      \item \textit{Notes:} Contribution of the sovereign risk shock to the forecast error variance at the given horizon. The forecast horizon is in months. The forecast error decomposition is based on the median of the impulse responses.
    \end{tablenotes}
  \end{threeparttable}

\end{table}

\Cref{tab:fevd} reports the relative contribution of the sovereign risk shock to the forecast error variance of the variables at various forecast horizons. The sovereign risk shock accounts for between 3 and 14 percent of the forecast error variance of output at the 24-month horizon. The forecast error variance of output increases over time, which points to the persistence of the sovereign risk shock. The estimates are similar to estimates by \cite{Bahaj2019}, who reports that the sovereign risk shock accounts for 15 percent of the forecast error variance of industrial output at a horizon greater than 18 months. Moreover, the sovereign risk shock accounts for approximately 6 to 14 percent of the forecast error variance of loan volumes at the two-year horizon.

\subsection{Sovereign risk shocks and government bond spreads}

As discussed in the previous section, the impulse response functions show that the sovereign risk shock affecting the financial sector tends to have a persistent effect on the government bond spread in each country. Next, we turn to discuss the contribution of the sovereign risk shock to the evolution of government bond spreads by carrying out a similar counterfactual exercise that eliminates the effect of the sovereign risk shock, as for output above. 

\begin{figure}
\caption{The difference between the actual government bond spread and the counterfactual in the absence of the sovereign risk shock}
\bigskip
\includegraphics[scale=0.9]{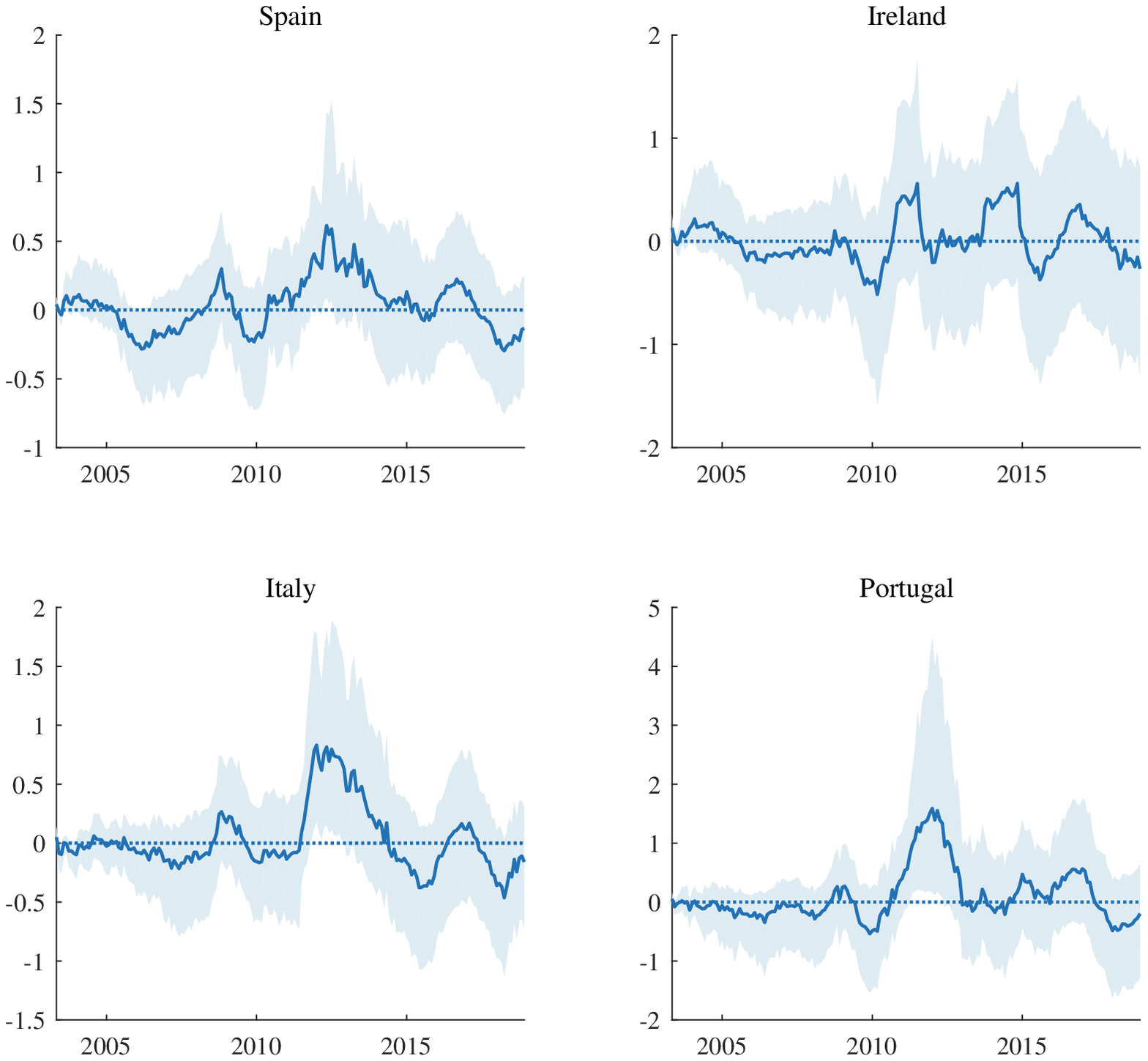}
\smallskip
\caption*{\textit{Notes}: The solid line is the median difference between the actual government bond spread and the counterfactual. Shaded areas mark the 68 \% credible intervals.}
\label{cf63}
\end{figure}

The differences between the actual and counterfactual time series are shown in \Cref{cf63}. The figure shows that the sovereign risk shock contributed to higher government bond spreads in all countries during the sovereign debt crisis, although the effect on the government bond spread in Ireland is more muted. In 2012, the sovereign risk shocks contributed to an increase in government bond spreads, with the median ranging from 75 basis points in Italy and Spain to approximately 150 basis points in Portugal. These values are in most cases below the estimates of the overall effect of unanticipated changes in government bond spreads in previous studies. \cite{Bahaj2019} reports that shocks to government bond spreads unrelated to economic conditions increased sovereign risk premiums by 100 basis points in Spain and by 150 basis points in Italy\footnote{\cite{Bahaj2019} reports the government bond spread of the two-year yield over the comparable German bond}. The reported increase is even larger for Ireland and Portugal where government bond spreads increased by more than 400 and 700 basis points, respectively. The estimates are also below those in \cite{Neri2013} who reported an increase in sovereign borrowing costs increased private borrowing costs in the crisis-hit countries by 1.3 percentage points on average. \Cref{tab:fevd} shows that the sovereign risk shock accounts for between 8 and 16 percent of forecast error variance of the government bond spread at 12 month-horizon. This suggests that a non-negligible share of the government bond spreads is explained by the sovereign risk shock that is transmitted via the banking sector, lending support to the transmission of bank risk to the sovereign. 

The results are consistent with the historical account of the evolution of the crisis, which began in Greece and Ireland and subsequently spread to the rest of the peripheral euro area countries. The results could be interpreted as evidence in favor of the sovereign risk largely passing through to the economy via the banking sector in all countries and the existence of a feedback loop between economic conditions and sovereign risk.

\subsection{Real effects of credit supply shocks}

\begin{figure}
\caption{Impulse response functions to a contractionary credit supply shock in the pooled panel SVAR model}
\bigskip
\includegraphics[scale=0.9]{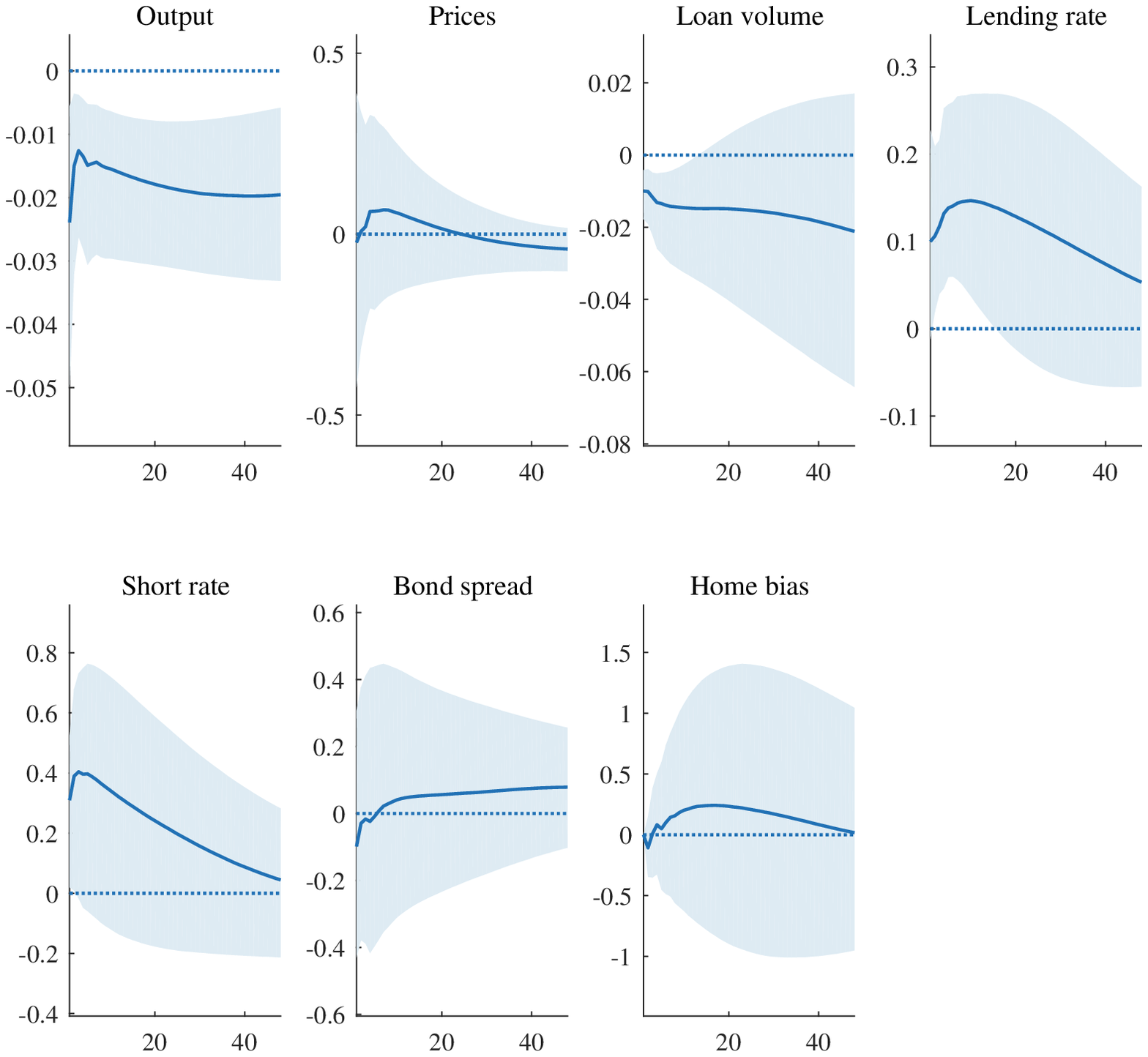}
\smallskip
\caption*{\textit{Notes}: The solid line is the posterior median and the shaded regions mark the 68 \% credible intervals. The shock is normalized, such that the lending rate increases by 10 basis points on impact.}
\label{irf_cs_pooled}
\end{figure}

\begin{sidewaysfigure}
\caption{Impulse response functions to a contractionary credit supply shock in the partially pooled panel SVAR}
\bigskip \bigskip
\includegraphics[scale=0.9]{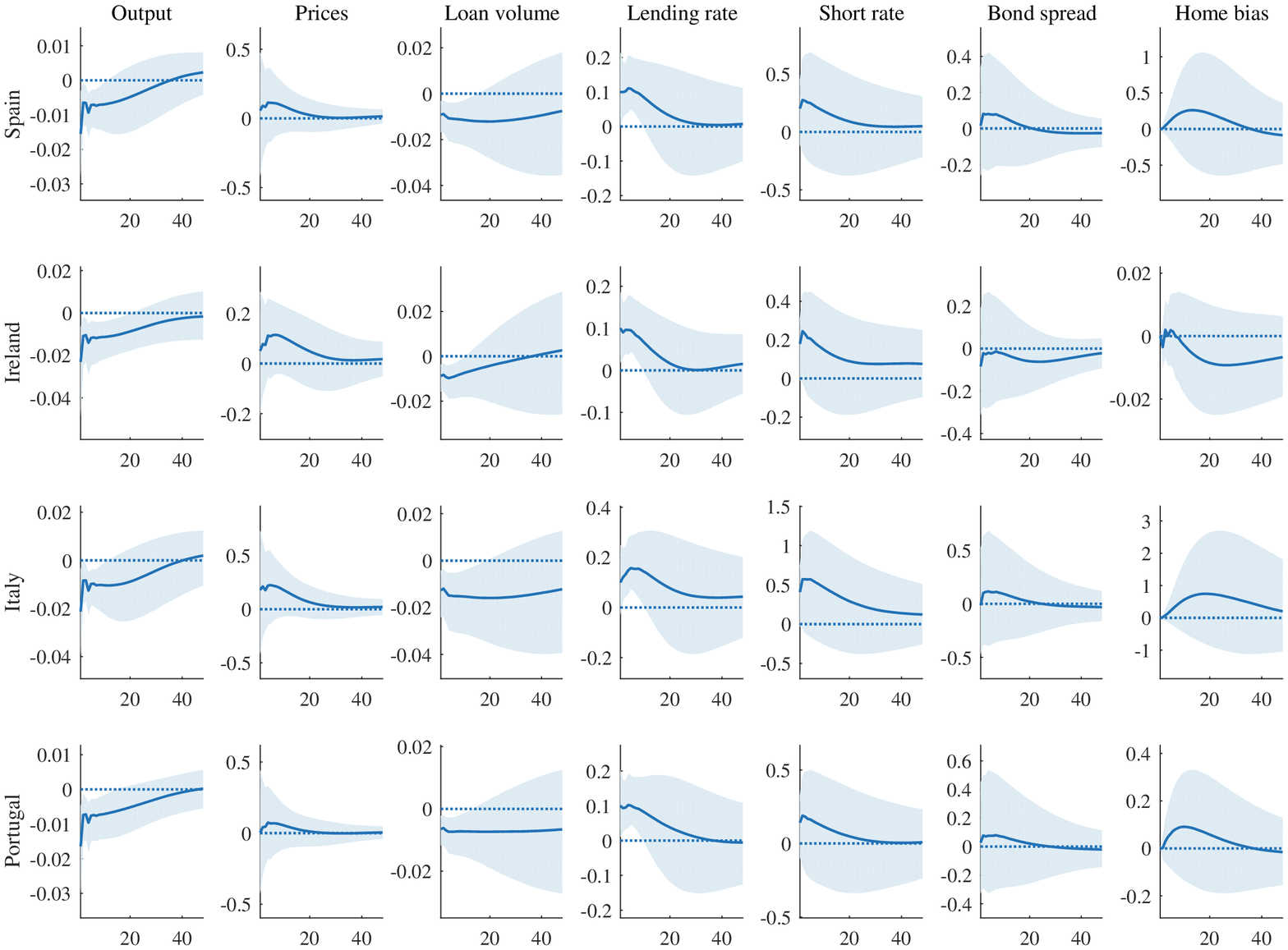}
\smallskip
\caption*{\textit{Notes}: See the notes in \Cref{irf_cs_pooled}}
\label{irf_cs}
\end{sidewaysfigure}

In this section, we discuss how the credit supply shock contributed to real economic activity in relation to the sovereign risk shock during the sovereign debt crisis. The impulse responses of the endogenous variables to the credit supply shock in the pooled and the partially pooled model are displayed in \Cref{irf_cs_pooled} and \Cref{irf_cs}, respectively. Each shock is normalized such that the median response of the retail lending rate is 10 basis points on impact. As \Cref{irf_cs_pooled} shows, a contractionary credit supply shock leads to a persistent decline in industrial output. The effect on output is negative with high posterior probability for around 10 months. The partially pooled model displays a less persistent effect of the credit supply shock on output, as is shown in \Cref{irf_cs}. The negative effect on output dissipates with high posterior probability within a year in each country. A contractionary credit supply shock also leads to an immediate and protracted decline in loan volumes; however, the long-run effect is not precisely estimated. The increase in the retail lending rate lasts with high posterior probability for around a year. Moreover, a credit supply shock does not have a significant effect on the government bond spread and is more likely to reduce the home bias of domestic banks.

\begin{figure}
\caption{The difference between the actual output and the counterfactual in the absence of the credit supply shock}
\bigskip
\includegraphics[scale=0.9]{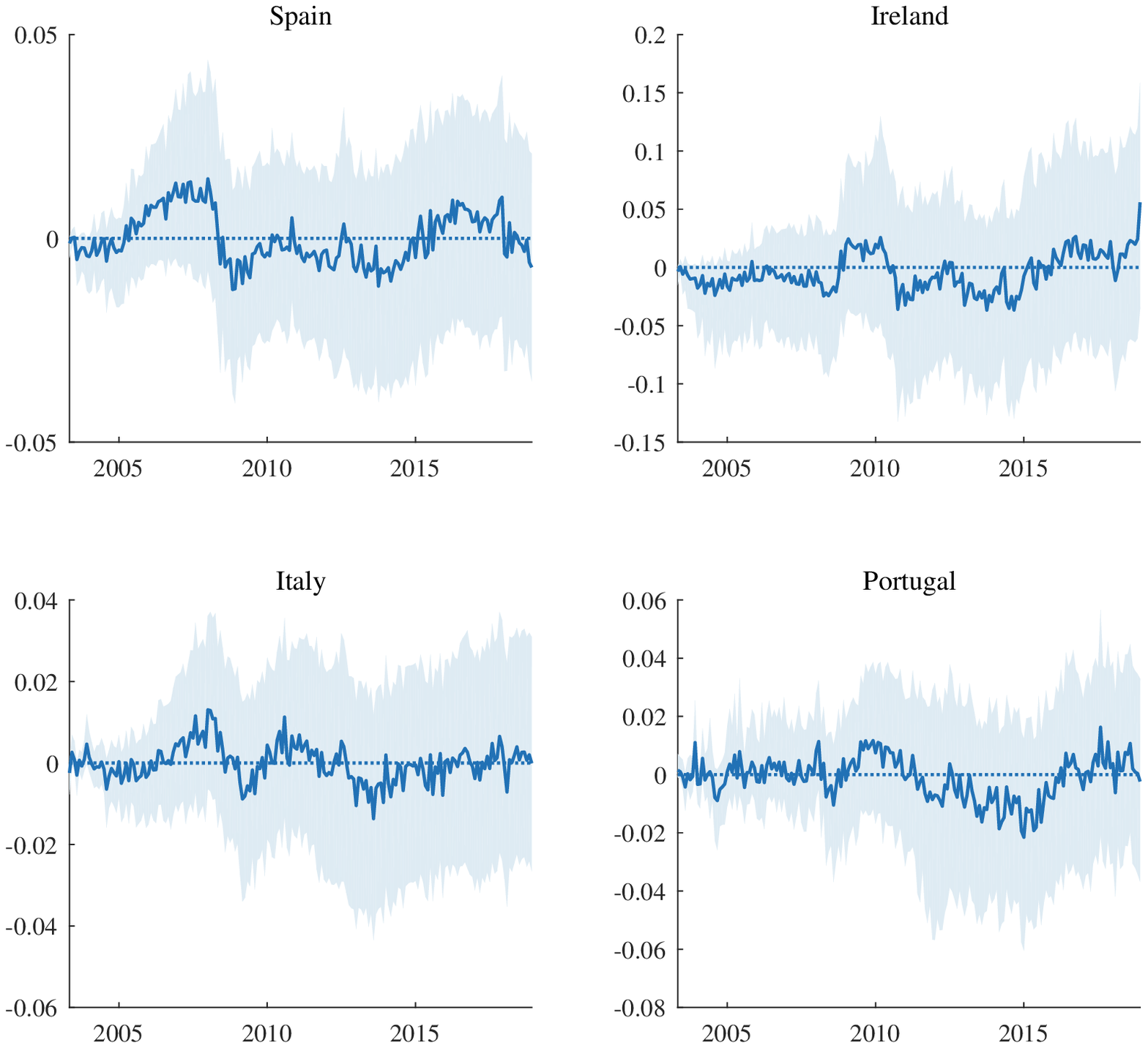}
\smallskip
\caption*{\textit{Notes}: The solid line is the median difference between the actual output and the counterfactual. Shaded areas mark the 68 \% credible intervals.}
\label{cf11}
\end{figure}

Although the estimates are not precise, the historical decompositions, shown in \Cref{cf11}, suggest that the credit supply shock had a positive median effect on economic output in Italy and Spain between 2006 and 2008 prior to the Global Financial Crisis. However, the countries were negatively affected to a varying degree by the disturbances in the supply of credit after the onset of the financial crisis in 2008. The contraction in output starting in 2008 was especially pronounced in Spain, where the beginning of the downturn coincided with the bursting of the Spanish property bubble. Moreover, the figure shows that credit supply shocks negatively affected real economic activity particularly in Portugal during around 2015. The results suggest that the sovereign risk shock accounts for a considerable share of the negative effect on economic output at the height of the crisis in 2012, but the credit supply shock also partly explains the decrease in output after the financial crisis.

\subsection{Discussion of results}

As discussed above, the results suggest that unanticipated changes in the sovereign risk reduced economic output in Italy, Portugal and Spain at the height of the crisis in 2012. This finding supports the arguments made in the literature that sovereign risk shocks affect real economic activity via the bank-lending channel primarily due to domestic banks' large exposure to sovereign debt \citep{Bocola2016} and the existence of a feedback loop between sovereign risk and the real economy \citep{Brunnermeier2016}. The results also corroborate the findings of recent empirical studies using micro-level \citep{Acharya2018, Bofondi2017, DeMarco2019, Gennaioli2018, Popov2014} and country-level \citep{Bahaj2019} data that document a tightening of financial conditions following an increase in government bond spreads during the European sovereign debt crisis. A notable exception is Ireland, where the economic output was not largely affected by the sovereign risk shock. This suggests that bank recapitalizations at the early stage of the crisis might have helped to break the vicious cycle between banks and sovereigns.

The results also suggest that the sovereign risk shock contributed positively to output via looser funding conditions in Spain, Portugal and Italy before the financial crisis. The positive effect of the credit supply shock on economic growth during the pre-crisis period is also documented by \cite{Bijsterbosch2015}. Moreover, in Portugal the negative effect of the sovereign risk shock largely dissipated after the government financial assistance program in 2012. 

These observations are important from a policy perspective in two ways. First, low government borrowing rates may have contributed to the overheating of the economy prior to the crisis, which suggests that tighter monetary policy could have helped to contain the economic boom. Second, government bailout programs and actions by the ECB to address financial market fragmentation appear to be effective in subduing the crisis by supporting the functioning of the financial sector. This suggests that policy reforms to prevent spillovers from the banking sector to the sovereign, and vice versa and actions to strengthen risk tolerance of the financial sector would play an important role in preventing future crises.

Although we do not explicitly disentangle the channels through which sovereign risk affects bank lending, the persistent increase in the domestic holdings of sovereign debt following increases in sovereign risk suggest that the reduction in bank lending in crisis-hit countries may be attributed to crowding out of corporate lending in favor of risky sovereign debt. \cite{Acharya2018} and \cite{Becker2017} also document changes in banks' asset composition in response to sovereign risk, which might be the results of banks' risk shifting or moral suasion by domestic regulators (see, e.g., discussion in \cite{Acharya2018}). In addition to crowding out lending, the increase in exposure to sovereign risk might have exacerbated the crisis. This finding suggests that the risk-weighting associated with banks' government bond holdings should be reassessed and implies that risky sovereign debt should be replaced by a European safe-asset as a store of value, as proposed by \cite{Brunnermeier2016}.

The results lend support to the view that credit supply shocks play a significant role in business cycle fluctuations in European countries (see, e.g., \cite{Hristov2012}). The results also show that credit supply shocks supported growth prior to the financial crisis in Italy, Portugal, and Spain, until they negatively contributed to output growth from 2008 to 2014.

\subsection{Alternative measure of real economic activity}
\label{sec:alt}
The results of the baseline model are largely unaffected by an alternative measure of real economic activity. To assess the robustness of the main results, we estimate the panel SVAR model using the unemployment rate for each country as the measure of economic activity. As in the baseline model, the VAR model includes a consumer price index, volume of loans to non-financial corporations, ratio of banks' holdings of domestic to foreign government debt, the retail lending rate on new loans, the government bond spread for each country and a proxy for the policy rate. The sign restrictions used to identify the three shocks are the same as in the baseline model (\Cref{tab:restrictions}), with the exception that the effects on the unemployment rate are assumed to be positive. The impulse response functions and the historical decompositions of the identified shocks are qualitatively similar to those obtained in the baseline models and do not alter the conclusion drawn from the baseline model.

\begin{sidewaysfigure}[htpb]
\caption{Impulse response functions to a contractionary sovereign risk shock in the alternative model}
\bigskip \bigskip
\includegraphics[scale=0.9]{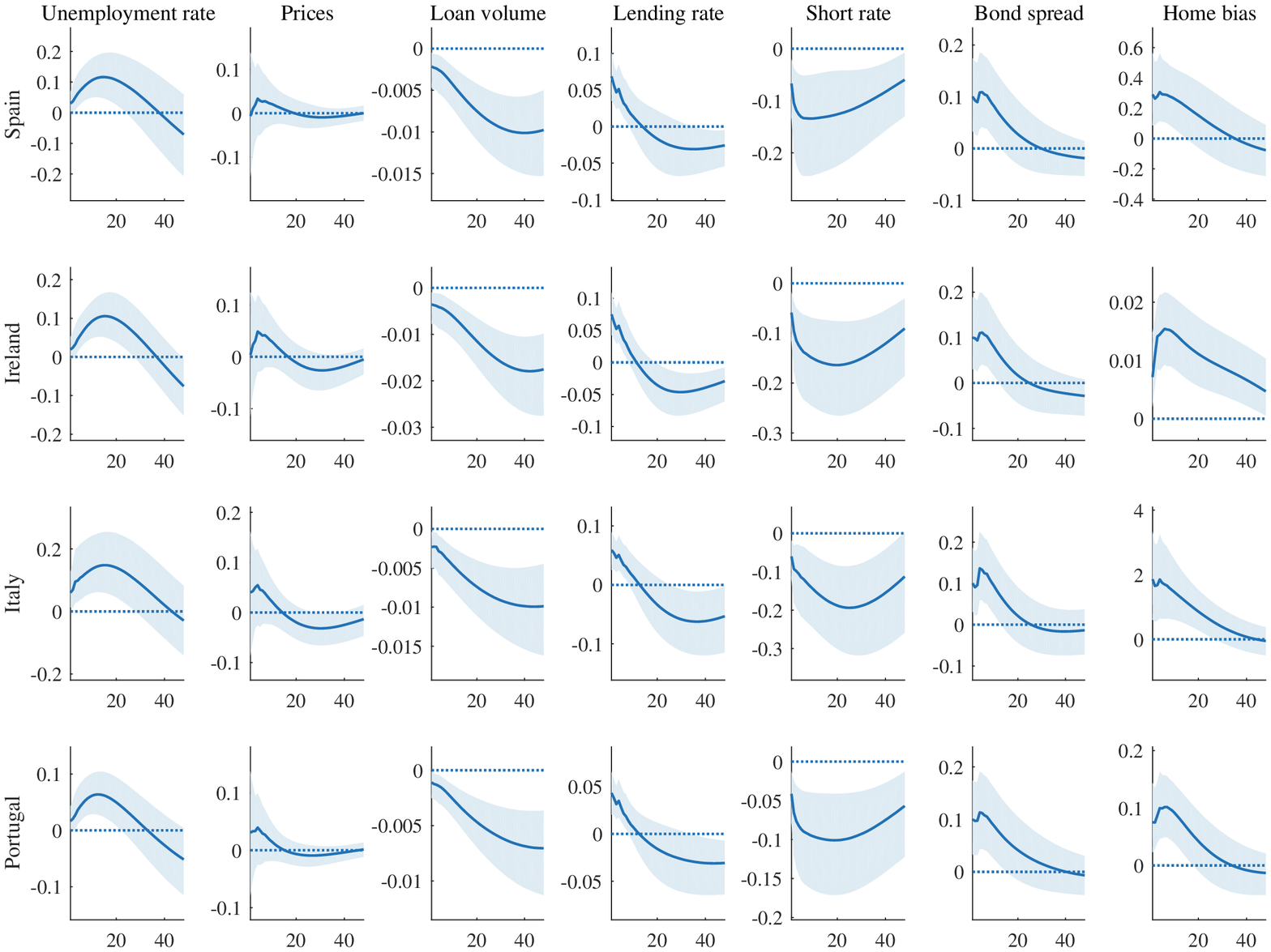}
\smallskip
\caption*{See the notes in \Cref{irf_sr_pooled}}
\label{irf_sr_u}
\end{sidewaysfigure}

\begin{sidewaysfigure}[htpb]
\caption{Impulse response functions to a contractionary credit supply shock in the alternative model}
\bigskip \bigskip
\includegraphics[scale=0.9]{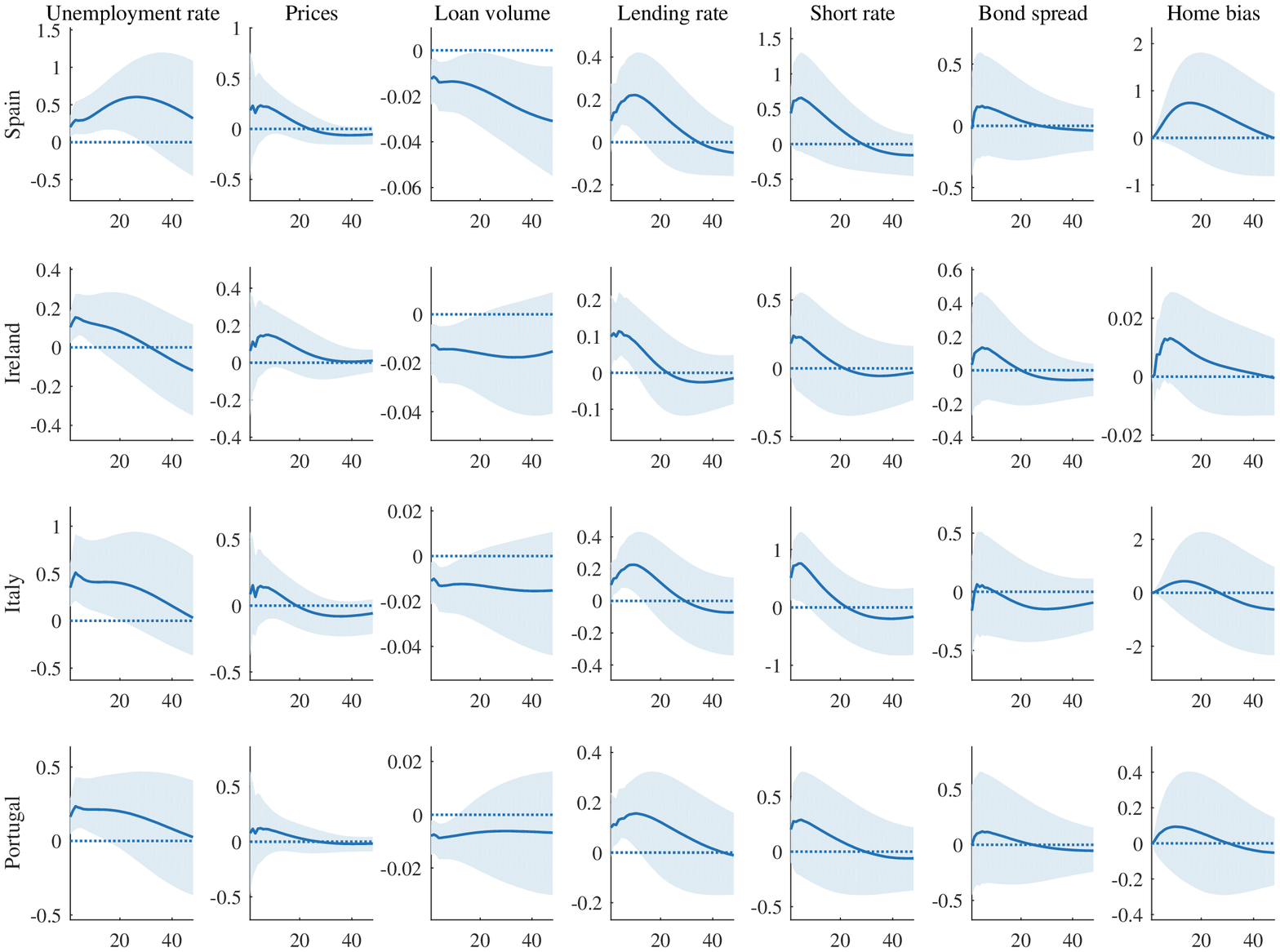}
\smallskip
\caption*{See the notes in \Cref{irf_sr_pooled}.}
\label{irf_cs_u}
\end{sidewaysfigure}

The impulse responses to the sovereign risk shock and the credit supply shock are presented in \Cref{irf_sr_u} and \Cref{irf_cs_u}, respectively. The impulse response functions to a contractionary sovereign risk shock show a large and persistent increase in the unemployment rate. As in the baseline model, an increase in sovereign risk corresponds with a persistent decrease in the volume of loans. Moreover, the immediate increase in the lending rate is likewise short-lived. The sovereign risk shock also has a persistent effect on the government bond spread, which is similar across the countries. Sovereign risk shocks affect the home bias of domestic banks in varying degrees across the countries. The effect of a sovereign risk shock increasing the government bond spread by 10 basis points on the home bias ranges from 1 basis point for Ireland to 200 basis points for Italy.

The impulse response functions to a contractionary credit supply shock (\cref{irf_cs_u}) also leads to a persistent increase in unemployment. The positive impact on unemployment dissipates after two years in the median response. The shock is also related to a persistent decrease in lending volumes across all countries. As in the case of the sovereign risk shock, the effect on the lending rate quickly dissipates following the shock. The impulse response functions also display a negligible effect on the government bond spread or the home bias of domestic banks.

\begin{figure}
\caption{The difference between the actual unemployment and the counterfactual in the absence of the sovereign risk shock in the alternative model}
\bigskip
\includegraphics[scale=0.9]{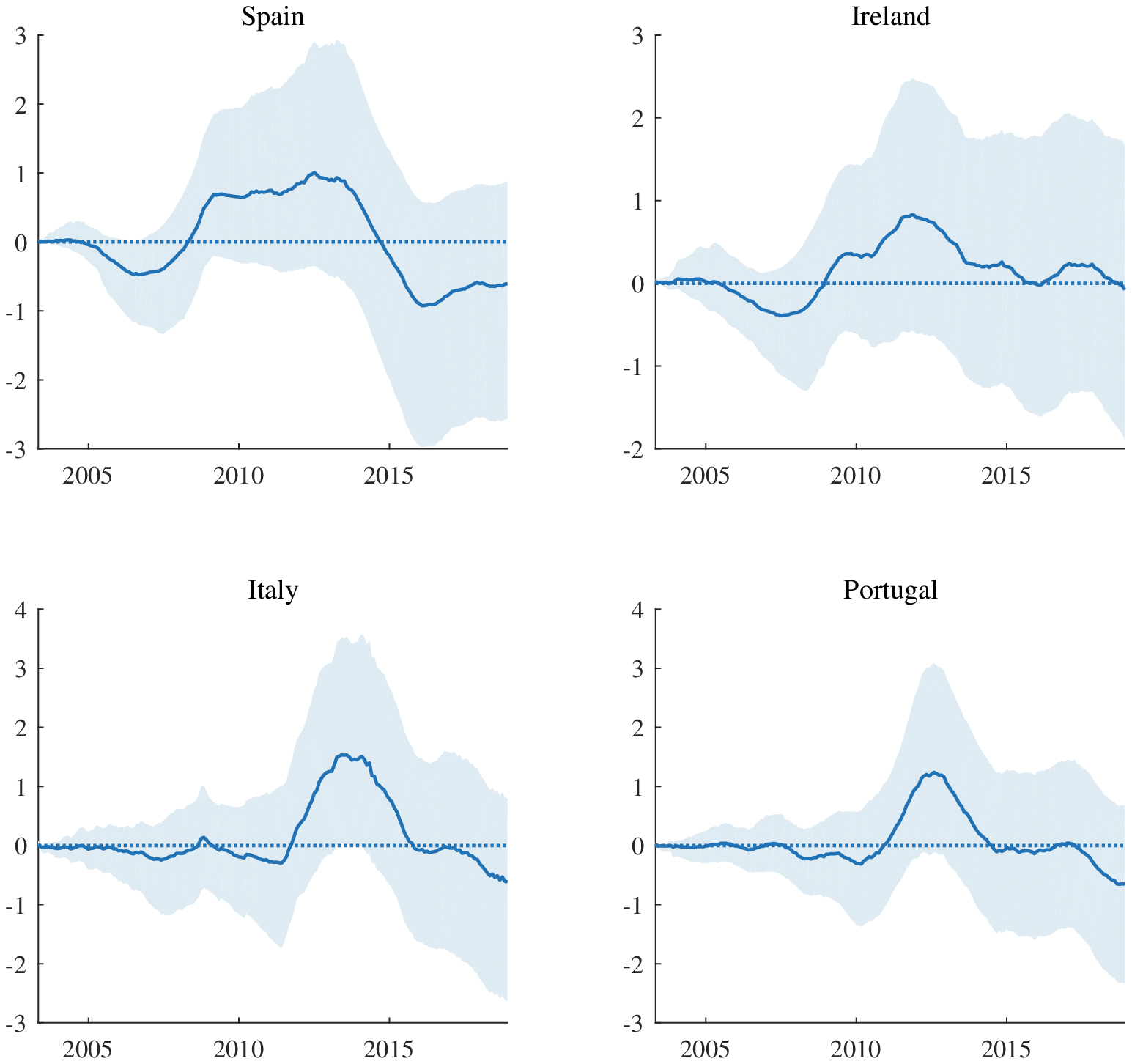}
\smallskip
\caption*{\textit{Notes}: The solid line is the median difference between the actual unemployment and the counterfactual. Shaded areas mark the 68 \% credible intervals.}
\label{cf13_u}
\end{figure}

The historical decompositions support the results from the baseline model. \cref{cf13_u} shows that sovereign risk shocks led to a median increase in unemployment between 2012 and 2015. Although the effects are not precisely estimated, the historical decompositions indicate that sovereign risk shocks led to a median increase of one percent in unemployment during the crisis. In comparison to the baseline results, sovereign risk shocks appear to have contributed to higher unemployment in Ireland. The difference between the baseline results might arise due to differences in the structure of the economy.

\section{Conclusion}\label{sec:conclusion}

The purpose of this paper is to re-examine the spillover of sovereign risk to the real economy via the banking sector. We proceed to evaluate the macroeconomic effects of sovereign risk shocks during the sovereign debt crisis by estimating a sign-restricted panel SVAR model for Italy, Ireland, Portugal and Spain. The results show that an increase in government bond spreads contributed to fall in economic activity in Portugal, Italy, and Spain and to a lesser extent in Ireland during the sovereign debt crisis. Moreover, the sovereign risk shock largely accounted for the disruptions in the credit supply during the debt crisis.

The findings in this paper point to several avenues for future research. First, identifying other transmission channels of sovereign risk would provide evidence of the relative importance of the bank-lending channel in the pass-through of sovereign risk in explaining fluctuations in real economic activity. To this end, estimation of SVAR models incorporating the housing market would be interesting to investigate the contribution of the real estate markets to bank and sovereign risk. Second, assessing the sensitivity of the transmission channel to banks' exposure to sovereign debt would be of interest from a macroprudential policy perspective.

\newpage

\section*{Acknowledgements}

I also gratefully acknowledge financial support from the Academy of Finland (grant no. 308628), the Yrj{\"o} Jahnsson Foundation (grant no. 6609, 7016), the Foundation for the Advancement of Finnish Securities Markets, and the Nordea Bank Foundation. I would like to thank Markku Lanne, Henri Nyberg, Joshua Aizenman and two anonymous referees for their helpful comments on the manuscript. I also thank the participants at the FDPE and Helsinki GSE Econometrics Workshops and the Freie Universit{\"a}t Berlin seminar on Topics in Time Series Econometrics for useful comments.

\bibliographystyle{apalike}
\bibliography{RHSbib}  

%
%
%
%

\clearpage

\setcounter{section}{0}
\setcounter{equation}{0}
\renewcommand{\theequation}{\thesection.\arabic{equation}}
\renewcommand{\thetable}{\thesection.\arabic{table}}

\setcounter{figure}{0}
\setcounter{table}{0}

\appendix
\section*{Appendix}
\renewcommand{\thesection}{A}

\subsection{Description of data}

A detailed description of the data is provided in \Cref{tab:data}.

\label{app:data}
\begin{sidewaystable}[h]
  \centering
  \tiny
  \caption{Description of data}
    \begin{tabularx}{\linewidth}{p{3cm}p{4cm}XXXXXX}
    \toprule
    \textbf{Name} & \multicolumn{1}{l}{\textbf{Description}} & \textbf{Unit} & \textbf{Frequency} & \textbf{Adjustment} & \textbf{Reference area} & \textbf{Code} & \textbf{Source} \\ \midrule 
    Industrial production & Volume index in production. Mining and quarrying; manufacturing; electricity, gas, steam and air conditioning supply & Index (2010=100) & Month & Seasonally and calendar adjusted & ES,IE,IT,PT & sts\_inpr\_m & Eurostat \\
    Unemployment rate & Unemployment rate of total population & Percentage & Month & Seasonally adjusted & ES,IE,IT,PT & une\_rt\_m & Eurostat \\
    Consumer prices & Harmonised index of consumer prices,  Overall index excluding energy and unprocessed food & Index (2015=100) & Month & No    & ES,IE,IT,PT & prc\_hicp\_midx & Eurostat \\
    Loan volumes & Outstanding amounts of adjusted loans to non-financial corporations in the euro area by monetary and financial institutions (MFI) excluding the ESCB & Millions (EUR) in the reference area & Month & No    & ES,IE,IT,PT & \multicolumn{1}{p{20.555em}}{BSI.M.ES.N.A.A20T.A.1.U2.2240.Z01.E\newline{}BSI.M.IE.N.A.A20T.A.1.U2.2240.Z01.E\newline{}BSI.M.IT.N.A.A20T.A.1.U2.2240.Z01.E\newline{}BSI.M.PT.N.A.A20T.A.1.U2.2240.Z01.E} & ECB \\
    Foreign government debt in banks' balance sheets & Outstanding amounts of general government debt securities of other Euro area member states (all countries except the reference area) held by MFIs in the reference area & Millions (EUR) & Month & No    & ES,IE,IT,PT & \multicolumn{1}{p{20.555em}}{BSI.M.ES.N.A.A30.A.1.U5.2100.Z01.E\newline{}BSI.M.IE.N.A.A30.A.1.U5.2100.Z01.E\newline{}BSI.M.IT.N.A.A30.A.1.U5.2100.Z01.E\newline{}BSI.M.PT.N.A.A30.A.1.U5.2100.Z01.E} & ECB \\
    Domestic sovereign debt in banks' balance sheets & Outstanding amounts of general government debt securities of the domestic country (the reference area) held by MFIs in the reference area & Millions (EUR) & Month & No    & ES,IE,IT,PT & \multicolumn{1}{p{20.555em}}{BSI.M.ES.N.A.A30.A.1.U6.2100.Z01.E\newline{}BSI.M.IE.N.A.A30.A.1.U6.2100.Z01.E\newline{}BSI.M.IT.N.A.A30.A.1.U6.2100.Z01.E\newline{}BSI.M.PT.N.A.A30.A.1.U6.2100.Z01.E} & ECB \\
     Cost of borrowing for corporations  & Annualised agreed rate on new loans to non-financial corporations, calculated by weighting the volumes with a moving average & Percent (p.a.) & Month & No    & ES,IE,IT,PT & \multicolumn{1}{p{20.555em}}{MIR.M.ES.B.A2I.AM.R.A.2240.EUR.N\newline{}MIR.M.IE.B.A2I.AM.R.A.2240.EUR.N\newline{}MIR.M.IT.B.A2I.AM.R.A.2240.EUR.N\newline{}MIR.M.PT.B.A2I.AM.R.A.2240.EUR.N} & ECB \\
    Government bond yield & Secondary market yields of government bonds with maturities of close to 10 years, monthly average & Percent (p.a.) & Month & No    & ES,IE,IT,PT & \multicolumn{1}{p{20.555em}}{IRS.M.ES.L.L40.CI.0000.EUR.N.Z\newline{}IRS.M.IE.L.L40.CI.0000.EUR.N.Z\newline{}IRS.M.IT.L.L40.CI.0000.EUR.N.Z\newline{}IRS.M.PT.L.L40.CI.0000.EUR.N.Z} & ECB \\
    Swap rate & Zero-coupon euro swap curve with maturity of 10 years, end of period & Percent (p.a.) & Month & No    & Euroarea & BBK01.WX0082 & Bundesbank \\
    Shadow Short Rate & Estimate for the euro area shadow short rate , month end& Percent (p.a.) & Month & No    & Euro area & & \cite{Krippner2013} \\ \bottomrule
    \end{tabularx}%
  \label{tab:data}%
\end{sidewaystable}%

\clearpage

\subsection{The sign and zero restriction algorithm}

\label{algorithm}
We proceed to identify the structural shocks using a mixture of sign and short-run exclusion restrictions using the method by \cite{Arias2018}. The reduced-form residuals in the VAR model in \cref{var} are related to the structural errors $\matr{\epsilon}_{c,t}$ by a linear transformation, such that 

\begin{equation*}
\matr{u}_{c,t} = \matr{P}_{c}\matr{\epsilon}_{c,t}.
\label{errors}
\end{equation*}

The structural errors $\matr{\epsilon}_{c,t}$ are orthogonal and normalized such that their variance-covariance matrix is $\mathbb{E}[\matr{\epsilon}_t \matr{\epsilon}_t'] \equiv \matr{\Sigma}_\epsilon = \matr{I}_K$ and hence $\mathbb{E}[\matr{u}_{c,t} \matr{u}_{c,t}'] \equiv \matr{\Sigma}_c = \matr{P}_{c}\matr{P}_{c}'$.

For any orthonormal matrix $\matr{Q}$\footnote{By definition, any  square orthonormal matrix $\matr{Q}$ satisfies $\matr{Q}\matr{Q}'=\matr{Q}'\matr{Q}=\matr{I}_N$.}, $\matr{\epsilon}_{c,t}^{*} = \matr{Q}\matr{\epsilon}_{c,t}$ is a candidate solution for structural shocks with unit variance and uncorrelated shocks. The candidate shocks are related to the reduced-form errors, such that $\matr{u}_{c,t} =  \matr{P}_{c}\matr{Q} \matr{\epsilon}_{c,t}^{*} $. The structural impulse responses $\matr{\Theta}_{c,i}^{*} = \frac{\partial \matr{y}_{c,t+i}}{\partial \matr{\epsilon}_{c,t}^{*}}$ for $i=1,\ldots,H$ may be obtained from $\matr{P}_{c}\matr{Q}$ and the reduced-form autoregression coefficients, and the candidate solution $\matr{\epsilon}_{c,t}^{*}$ is retained if the shocks satisfy the specified restrictions on the impulse response functions, as specified in \Cref{sec:identification}. 

Let $\matr{\Theta}_{c}^{*}$ denote a matrix containing all of the impulse responses for periods $h = h_1, \ldots, h_n$ for which the restrictions are imposed, such that

\begin{equation*}
\matr{\Theta}_{c}^{*} = \begin{bmatrix}
\matr{\Theta}_{c, h_1}^{*} \\ \vdots \\ \matr{\Theta}_{c, h_n}^{*}
\end{bmatrix}.
\end{equation*}

Sign and zero restrictions can be represented by selection matrices $S_j$ and $E_j$ for $j=1,\ldots, K$, respectively, where the number of columns are equal to the number of rows in $\matr{\Theta}_{c}^{*}$. The number of rows in $S_j$ is equal to the number of sign restrictions imposed on each shock $j$, and each row represents a single non-zero restriction on the impulse responses.

The model parameters satisfy the sign restrictions if $S_j \matr{\Theta}_{c}^{*(j)} >0$ for all $j=1,\ldots,K$, where $\matr{\Theta}_{c}^{*(j)}$ represents the $j$-th column of the matrix $\matr{\Theta}_{c}^{*}$. For zero restrictions, $E_j$ is a selection matrix where a single non-zero element in each row specifies a zero restriction and the number of rows represent the number of zero restrictions for each shock $j$. The zero restrictions are satisfied if $E_j \matr{\Theta}_{c}^{*(j)} = 0$ for all $j=1,\ldots,K$.

The impulse response functions are obtained using the algorithm of \cite{Arias2018} that combines sign and exclusion restrictions. 

\begin{enumerate}
\item  Draw $\matr{\Sigma}_{c,u}$ and $\matr{A}{c}$ from the posterior of the model parameters. Applying a Cholesky decomposition to $\matr{\Sigma}_{c,u}$, we obtain the lower-triangular matrix $\matr{P}_{c}$ that satisfies $\matr{P}_{c}\matr{P}_{c}^{'}=\matr{\Sigma}_{c,u}$. 

\item Draw an orthogonal matrix $\matr{Q}$ that satisfies the zero restriction $E_j \matr{\Theta}_{c}^{*(j)} = 0$ for $j=1,\ldots,K$ by drawing each column of a $K \times K$ matrix $\matr{W}$ from a $N(\matr{0},\matr{I}_K)$ distribution and applying a QR-decomposition to each draw of $\matr{W}$.

\item Calculate impulse responses $\matr{\Theta}_{c}^{*}$ using $\matr{Q}\matr{\Theta}_{c, 0}^{*}$. Retain draw if $S_j \matr{\Theta}_{c}^{*(j)} >0$ for all $j=1,\ldots,K$ else return to step 2. 

\item Repeat steps 1-3 until $B$ successful draws are obtained.

\end{enumerate}

\cite{Arias2018} specify an algorithm for obtaining $\matr{Q}$ in step 2. For $j=1,\ldots,K$, let
\begin{equation*}
R_j = \begin{bmatrix}
E_j\matr{\Theta}_{c}^{*} \\ Q_{j-1}
\end{bmatrix}
\end{equation*}

where $Q_{j-1} = [q_1,\ldots,q_{j-1}]'$ and $q_{j}$ is the $j$-th column of $\matr{Q}$. For $j=1$, $Q_{0}=0$, such that $R_1 = E_1\matr{\Theta}_{c}^{*} $. 

The algorithm to obtain $\matr{Q}$ is as follows:

\begin{enumerate}
\item Set $j=1$
\item Create $R_j$ and find the matrix $\matr{N}_{j-1}$ that  forms an orthonormal basis for the null space of $R_j$. 
\item Draw a $K \times 1$ vector $x_j$ from a standard normal distribution on $\mathbb{R}^K$.
\item Set $q_j = \matr{N_{j-1}}\left( \matr{N_{j-1}}^{'}x_j  / \norm{\matr{N_{j-1}}^{'}x_j}  \right)$ 
\item Set $j=j+1$ and move to step 2. If $j=K$ form matrix $\matr{Q}= [q_1,\ldots,q_K]$
\end{enumerate}

The algorithm allows us to impose at most $K-j$ zero restrictions for each shock $j$. If the number of restrictions for shock $j$ exceeds $K-j$, $\matr{Q}$ cannot be orthonormal, given that the orthonormal basis for the null space of $R_j$ would be trivial, such that $q_j=0$.

\end{document}